# LOW-COST SPECTRUM ANALYZERS FOR CHANNEL ALLOCATION IN WIRELESS NETWORKS 2.4 GHZ RANGE

*DSc Buryachok V. L.,*
*MSc Sokolov V. Yu.*

*Ukraine, Kyiv, State University of Telecommunications, Dept. of Information and Cyber Security*

**Abstract.** *The article introduces a new scheme of dynamic interference free channel allocation. The scheme is based on additional spectral analyzers in wireless networks IEEE 802.11. Design and implementation is presented.*

**Introduction.** The development and widespread wireless technologies leads to constant growth in the number of users and devices. But increasing of user's number in a limited frequency range and spatial leads to channel interference that ultimately impacts the bandwidth of wireless channels and even the overall performance.

According to the analysis of statistics from 2001 to 2016 collected by Wireless Geographic Logging Engine [1], the growth of the number of wireless access points (APs) in the world can trace the dynamics of growth in the number of networks and predict their number in the near future. Fig. 1 shows statistics for the growth of the number of wireless APs and its model.

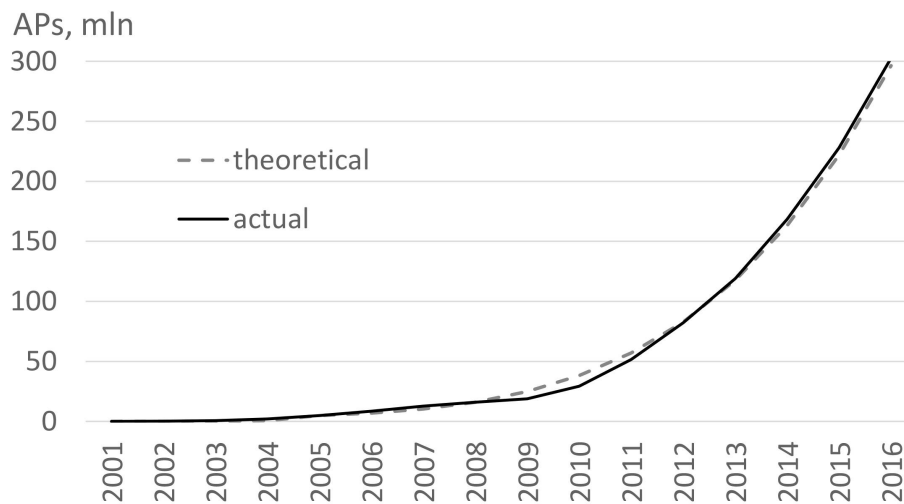

*Fig. 1. Growth in the number of wireless access points*

Power function is used to describe the process of growth of AP's number at the end of the year:

$$N(year) = (year - 2000)^{4,54} + 3450 \qquad (1)$$

where *year*—accounting year. Extrapolating from the graph and using the formula (1), at the beginning 2018 can be expected nearly 400 mln APs.

Dotted line on Fig. 1 shows the period of doubling the number of wireless networks (for the IEEE 802.11 standard). The average time of the update is less than two years and is consistent with Moore's Law. The next doubling of the amount of wireless networks is expected at the beginning of 2020.

A separate indicator is the period of doubling the number of wireless networks (for the IEEE 802.11 standard). From the graph shown on Fig. 2, shows that the average time of the update is less than two years and is consistent with Moore's Law. On the schedule for the crisis of 2008 has to a delay in almost three years, which is then compensated by the rapid growth in the number of APs. If the average growth rate of the number of (the graph indicated by the dotted line), we see that the process is slowing down, but it is rapid.

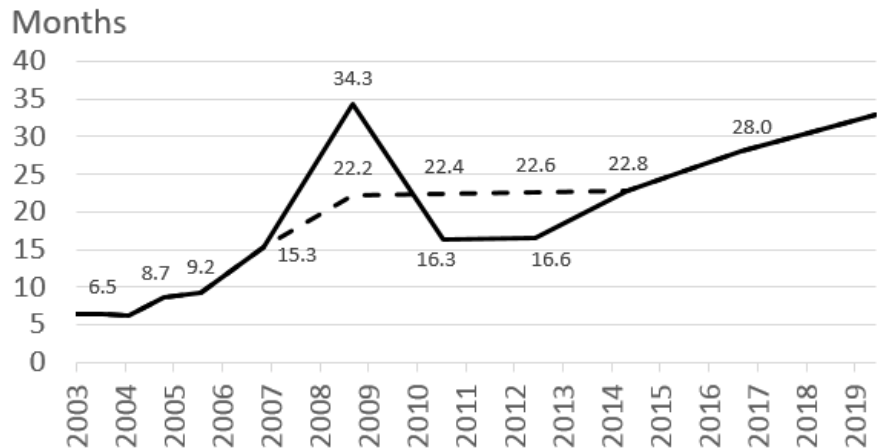

*Fig. 2. Terms of doubling the number of wireless APs*

In short, the expected exponential growth of APs demands improved channel allocation to ensure maintaining quality of service. The reminder of the article is organized as follows. In section "Overview of Methods of Improving Efficiency of Allocation of Frequency Resources" addresses different antenna technologies. Section "Methods of Spectrum Analysis" gives an overview of methods and related hardware. Our suggestion on improving channel allocations is given in section "Scheme of Systems with Spectrum Analyzers". The corresponding algorithm is given in section "Algorithm". Design and implementation of our algorithm is given in section "Implementation". The article end with section "Conclusion and Future Work and References".

**Overview of Methods of Improving Efficiency of Allocation of Frequency Resources.** The growth of wireless networks especially in dense of urban areas leads to a mutual influence on each other's networks. Moreover, the number of mobile devices and embedded systems that may be used as a wireless AP (e. g., the family of ESP8266 or RTL8710 modules) also increases. To solve the problem of effective usage of frequency resources can be applied different methods:

1. Administrative: centralized planning of wireless infrastructure, legal restrictions on the level of the signal or power transmitters.

2. Regular monitoring and adaptation of the systems manually. For example, using of the focusing antennas [2].

3. Adaptive tuning systems (dynamic channel allocation) at the protocol level (IEEE 802.11f, IEEE 802.11k), monitoring and auto-tuning at the level of the receive path, some manufacturers (Atheros Spectral Scan mode), use additional devices to gather information on the state of the wireless system.

These methods often do not take into account the influence of other wireless technologies that operate at the same frequency range, for example, by the standards of IEEE 802.15.1 (Bluetooth), IEEE 802.15.4 (ZigBee, MiWi, WirelessHART, ISA100.11) and other non-standard devices, as well as household and industrial noise.

Many manufacturers of wireless equipment have embedded dynamic channel allocation algorithms but the spectrum is scanned only near AP, thus, are not considered particularly customer location [3; 4]. AP starts at the clear channel at the place of its location that improves the performance of the entire network but does not make it optimal (as it is not possible to take into account all the parameters in ad-hoc networks: *polarization*, the height of the *shielding* and *reflections*, as well as the *movement* of the user). Because information from the AP about frequency setting user location (all or selective) should be considered also, and on the basis of the received data the best frequency channel should be selected. To collect the information you can use existing wireless cards but their range of visibility is often limited only by the IEEE 802.11 standard networks (and some maps do not even see "hidden" networks). That's why we propose to use additional independent devices as spectrum analyzers to improve performance of universal method of channel allocation.

**Methods of Spectral Analysis.** After comparing devices available on the market shows that the various chips with different quality to cope with the task of analyzing the spectrum. But also we found some spectrum analyzer's limitations such as:

1. They can be positioned stationary at key points in the infrastructure user locations.
2. The antenna pattern should be close to spherical (non-directional antenna).
3. To send the results to be used wireless or wired communication links, which go beyond the scanned frequency band.
4. The minimum power consumption (for off-line power capabilities).

Table 1 lists the most common chip that can be used as a signal for monitoring devices.

Table 1. Main Characteristics of Receivers

| Chip | Frequency range, MHz | Resolution, kHz | Range, dBmW | Sensitivity, dBmW |
|---|---|---|---|---|
| Nordic nRF24L01 | 2400–2525 | 977 | from –85 to –42 | 1.0 |
| Cypress CYRF6934 | 2400–2483 | 1000 | from –90 to –40 | ~4.1 |
| Cypress CYRF6935 |  |  | from –95 to –40 | ~3.1 |
| Cypress CYRF6936 | 2400–2497 |  | from –97 to –47 | ~1.3 |
| Chipcon CC2500 | 2400–2484 | 58–812 | from –104 to –13 | 0.8 |
| Chipcon CC2511-F32 |  |  | from –110 to –6.5 | 0.5 |

In addition, based on the same chip set portable spectrum analyzers:
− Cypress CYRF6934—MetaGeek Wi-Spy 2.4i.
− Cypress CYRF6936—Wi-Detector (ver. 2 and 3).
− Chipcon CC2511-F32—Pololu Wixel, Ubiquiti AirView2 and MetaGeek Wi-Spy 2.4x.

When we use chips from different vendors, we have to harmonize data that received from them. As used chips from multiple vendors, then use them in a single system should be harmonized received from them a result. The following Table 2 provides guidelines for quantitative gradation signal quality, the unit of measurement is taken dBm.

**Scheme of the System with Spectrum Analyzers.** We have developed our own system with the ability to obtain data from a variety of devices. The system consists of AP (or in some systems with roaming) customers and spectrum analyzers, which are connected to the controller or to the client via wired links (with PoE).

The scheme shown on Fig. 3 may or may not contain a spectrum analyzer on the AP side. The dotted line displays the spectrum analyzer that is either connected to AP or is a part of AP.

Table 2. Main Characteristics of Receivers

| Signal level, dBm | Quality | Specifications |
|---|---|---|
| less than –90 | Unacceptable | Close to the intrinsic noise of equipment and background. Any work unlikely |
| from –90 to –81 | Bad | The minimum signal level to maintain network integrity. Shipping packages can be unreliable |
| from –80 to –71 | Acceptable | The minimum signal level for reliable packet delivery, such as email, web |
| from –70 to –67 | Very good | Signal strength for applications requiring very reliable, timely delivery of data packets, eg., VoIP, streaming video |
| more than –67 | Excellent | The maximum achievable signal. The client is a few meters a few feet from the transmitter. Atypical situation |

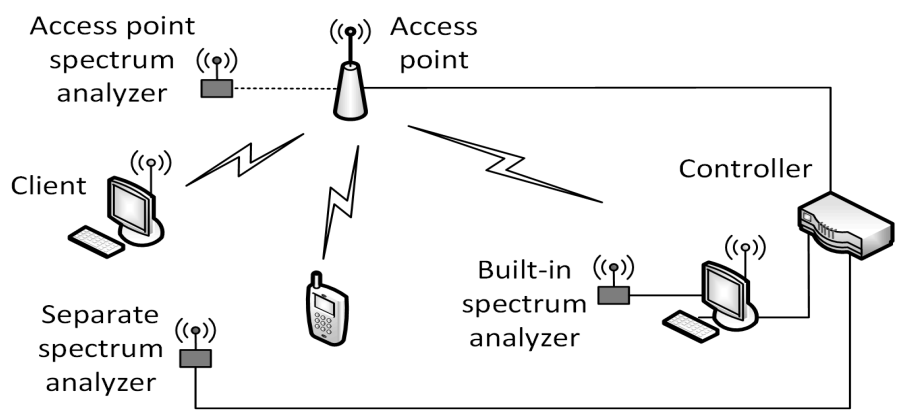

*Fig. 3. Scheme with spectrum analyzers*

The scheme consists of:
− *Controller* that distributes the channels and the load on the network, managing the wireless infrastructure.
− *Access point* that switches to silent mode, data is collected from available spectrum analyzers or internal wireless interface. Collected data sends to the controller. APs may be multiple.
− *Client* that collects data from its own wireless network card or from *built-in spectrum analyzer* and sends it to the controller.
− *Separate spectrum analyzer* that has Ethernet interface and send data directly to the controller.

All spectrum analyzers are assigned with weights based on the importance of their location.

Controller selects the free channel for each AP of the network and initialize transmission on the new channel. The scanning process is repeated.

The system can contain several segments that are separated by channels and/or physically separated. This approach makes it easy to scale the system with roaming.

The average value of the signal level in the *j* channel:

$$L_j^{ch} = \frac{1}{N} \sum_{i=1}^{N} L_{ij},\qquad(2)$$

where index *ch*—channel number; *N*—number of points that belong to the same channel; $L_{ij}$—measurement in the $i^{th}$ point for the $j^{th}$ channel, dBmW. For one measurement cycle is recommended to poll every point of about 100 times, because instead of $L_i$ is better to use an average value for the number of measurements.

The average signal from all j external devices:

$$\overline{L}_{ext}^{ch} = \frac{1}{M} \sum_{j=1}^{M} \mu_j L_j^{ch},\qquad(3)$$

where *M*—amount of spectrum analyzers; $\mu_j$—weighting the importance of a particular device; $L_j^{ch}$—average signal level for a particular device from (2).

The average signal from built-in wireless card is only received through the channel levels (and only for devices that work in the same standard):

$$\overline{L}_{int}^{ch} = \frac{1}{H} \sum_{j=1}^{H} \mu_j \sum_{k=1}^{K} v_k^{ch} L_k,\qquad(4)$$

where *H*—amount of built-in wireless cards; *K*—number of scanned wireless APs; $v_k^{ch}$—crossing channels coefficient due to channel width of 20 MHz and channel spacing—5 MHz (see Table 2); $L_k$—signal levels to $k^{th}$ wireless network.

**Algorithm of Dynamic Channel Allocation.** The controller receives information of two types: from spectrum analyzers and from network cards. The format of the input data is different, but for the final result it is enough to transfer to the controller a list of channels with a minimum signal level.

Therefore, the data from spectrum analyzers and from network cards are averaged and unified. Data is transmitted asynchronously, and the controller collects data, determines the emptiest channels for each AP, and generates a task for APs to change the channel. The algorithm is repeated. The Fig. 4 shows the data collection algorithm.

**Implementation.** Consider the three embodiments of portable spectrum analyzers:
1. The receiver and the control unit (separated and on a single chip).
2. The receiver with determined signal level on two states and the control unit.
3. The integrated client's wireless cards.

Since different types of equipment with different operating systems can be present in the system, with different speed and with different measurement accuracy, unification should be carried out at the information gathering stage.

Here are three types of spectrum analyzers:
1. With separated receiver and control module with USB.
2. With integrated receiver and control module with USB.

3. With separated receiver and control module with Ethernet.

Table 3. Crossing Channels Coefficient

| \|ch–k\| | 0 | 1 | 2 | 3 | >4 |
|---|---|---|---|---|---|
| $v_k^{ch}$ | 1 | ¾ | ½ | ¼ | 0 |

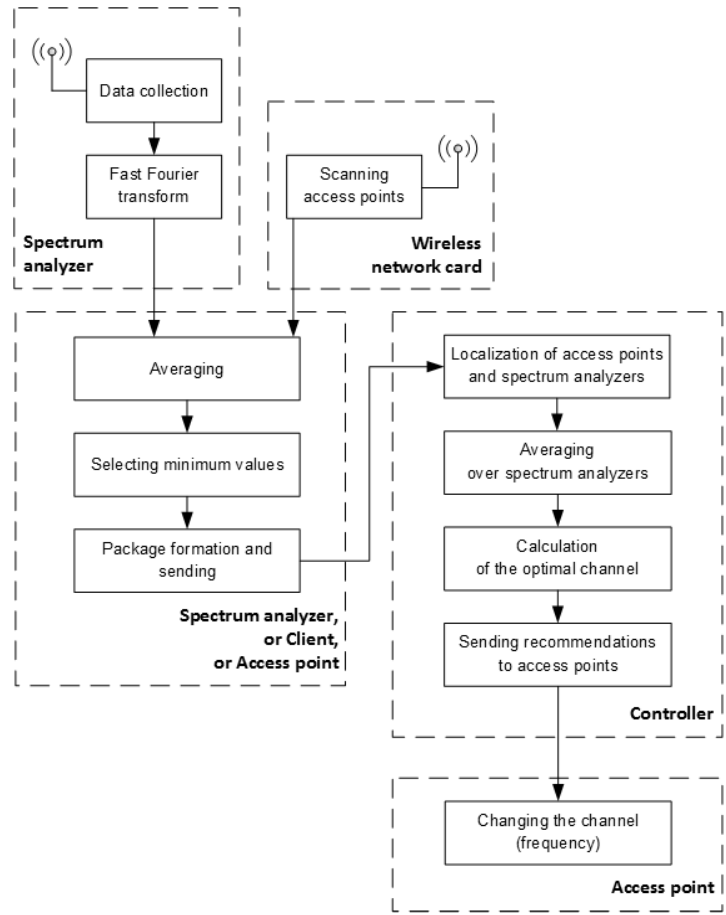

*Fig. 4. Algorithm of dynamic channel allocation*

The first two types of spectrum analyzers can work together with the client, with an AP or separately. The third type is intended for use separately.

Let's look at the implementation of a spectrum analyzer for separate scheme. The control unit implemented on Arduino Nano v. 3.0 (with 3.3V supply) and the receiver—on TI CC2500+PA+LNA module with external antenna [6]. Scheme is equipped with OLEDs 0.96" 128×64 SSD1306 (via I2C or SPI) for visualizing the instantaneous value. The program is written in Arduino IDE and compiled with GCC.

Fig. 5 and 6 show a diagram of the connection module and power to the device. This device can be operated with a controller that receives data via a USB interface. On two screens displayed range from 2400.01 to 2503.40 MHz with spacing in 405.5 kHz. It was found that the control unit memory is not enough (only 2 KB of RAM) to analyze the available channels. In addition, this device does not fulfill one of the requirements—must be non-directional antenna, and the half-wave dipole has a distinct polarization.

To implement the spectrum analyzer on a single chip has been selected Pololu Wixel, whose RAM size is 4 KB, the non-directional antenna, five times lower power consumption, almost one and a half times better resolution and SDK with detailed documentation [7]. The program is compiled with SDCC [8].

Fig. 6 and 7 show a diagram of the connection module and power to the device. On two screens displayed range of 2403.47–2476.50 MHz with spacing in 286.4 kHz, and information about the recommended Wi-Fi and ZigBee channels, calculated using the formula (3). The Fig. 8 shows an example of display device assembly in a transparent case with additional switcher for pause mode.

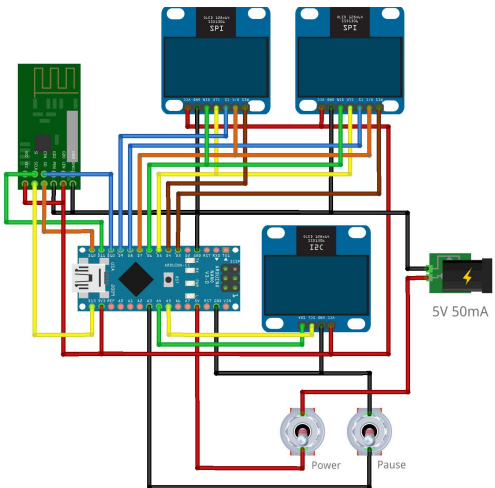

*Fig. 5. Schematic diagram of separated receiver and control unit*

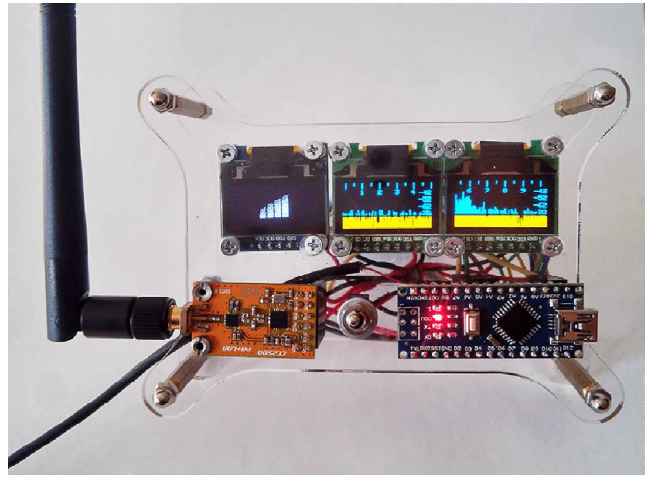

*Fig. 6. Separated receiver with external antenna and control unit*

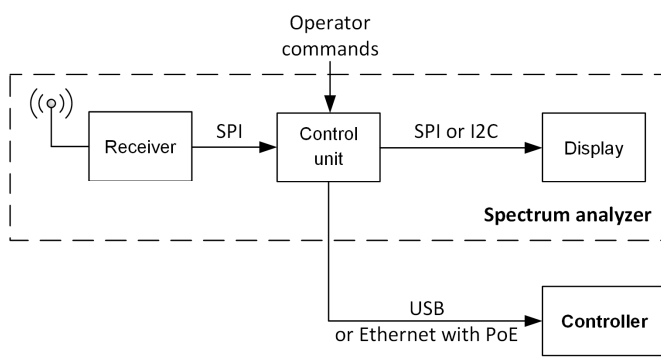

*Fig. 7. Schematic diagram of separated receiver and control unit*

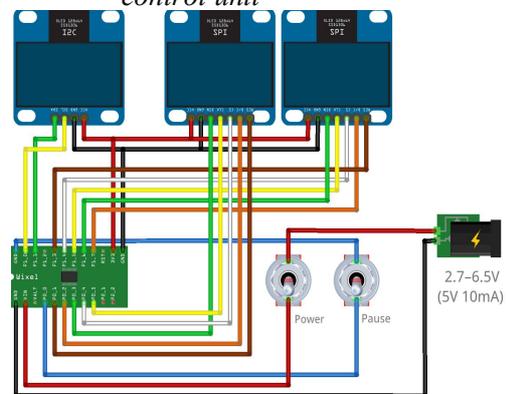

*Fig. 8. Schematic diagram of receiver and control unit on a single chip*

It can also be used to evaluate the signal strength and the module in which the signal level is determined by the evaluation of indirect methods. On nRF24L01 module (or its modification with power and low-noise amplifiers) base in Raspbian operating system for embedded systems platform Raspberry Pi. Chosen as the prototype of the spectrum analyzer, which was adapted for the microcontroller Broadcom BCM2837 SoC (core ARM Cortex A53 CPU) core with flags -march = armv8-a + crc -mtune = cortex-a53 -mfpu = neon-fp-armv8. To initialize the device in the operating system using third-party software RF24 library.

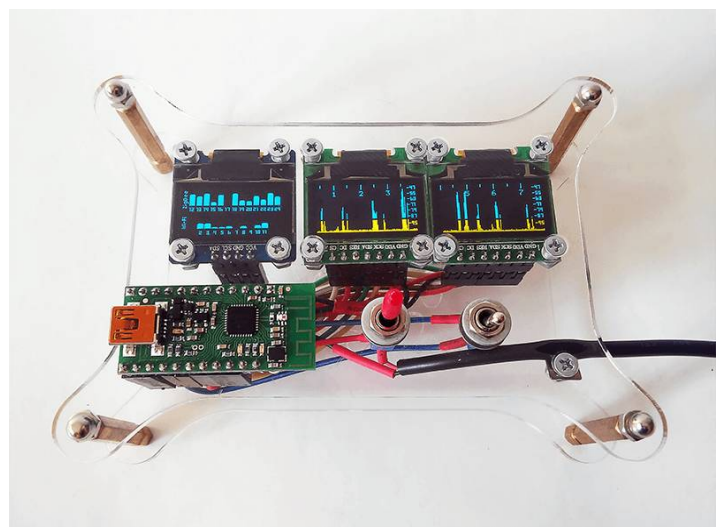

*Fig. 9. Receiver and control unit on a single chip*

The peculiarity of receiving data from nRF24L01 chip is the fact that it is only one flag (_NRF24_RPD), indicating that the level of the received signal is above or below level of –64 dBmW. Change the mode of the application, taking into account changes in the library bcm2835 (v. 1.50) [2]. The scanning is performed every 976.5625 kHz, thus covered a range of 2.400–2.525 GHz (128 measuring points).

As for the frequency of the receiver sensitivity of –85 dBmW [9], when you receive 200 measurements, the signal level in the $i^{th}$ point is calculated using the formula:

$$P_i = L_{\min} + \frac{2(L_{av} - L_{\min})}{N} \sum_{j=1}^{N} p_{ij}, \quad (5)$$

where $L_{min}$—minimum level, dBmW; $L_{av}$—flag trigger level, dBmW; $N$—number of measurements; $p_{ij}$—result of the single measurement (can take two values: 0 or 1).

From (5) we have a special case for this dimension:

$$P_i = -85 + 0.21 \sum_{j=1}^{200} p_{ij}. \quad (6)$$

The results implementing the calculation for measuring points are displayed on screen. In our case we had 128 measuring points and we will see an array of 128 values in command line.

A full-scale model was collected on the module, which is shown in Fig. 10. Arduino Nano v3 is used as the control system, and the data is output on the OLED SSD1306. To speed up processing, the process can be parallelized.

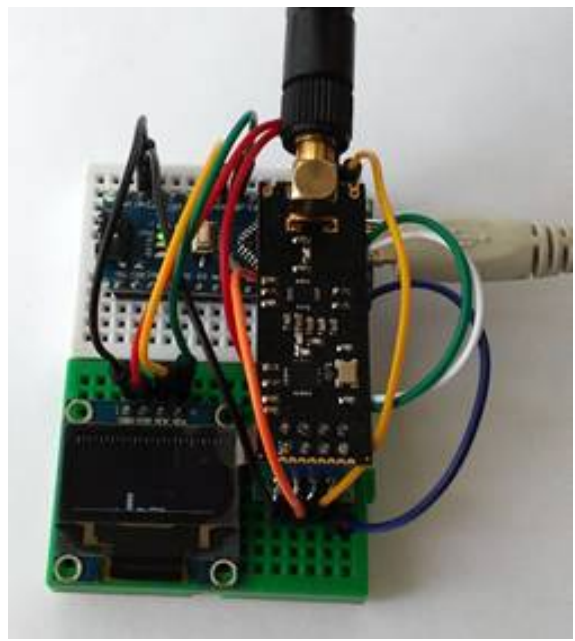

*Fig. 10. Receiver nRF24L01 with Arduino Nano control unit*

**Conclusion and Future Work.** It is proposed that use of a set of built-in spectrum analyzer, and wireless card on user side has potential to reduce interchannel interference. Spectrum analyzer prototypes has also been designed from industrial modules. For the implementation of end devices it is recommended to create a separate design with a protective casing, additional connector for Ethernet connection with PoE. Assessments and validations of the design as well as conducting validating experiments are planned in our IoT test environment.

This scheme does not require the use of a cloud. All data is transmitted over the internal network. In the event that the security policy allows access to the network of guests, the data should be encrypted using a block cipher. And since the size of messages is small, then the requirements for the stability of the algorithm and the length of the key are not high.